\documentclass[a4paper,12pt,svgnames]{article}
\usepackage[style=philosophy-classic,natbib=true]{biblatex}
\bibliography{manuscrito.bib}
\usepackage{graphicx,csquotes,amssymb,enumerate,amsmath,amssymb,mathtools,amsfonts,epigraph,setspace,mathrsfs,xcolor,ebgaramond-maths}
\usepackage[blocks]{authblk}
\usepackage[english]{babel}
\usepackage[colorlinks,allcolors=NavyBlue]{hyperref}
\title{\bf Towards a process-based approach to consciousness and collapse in quantum mechanics}

\date{Forthcoming in \textit{Manuscrito} vol. 47, n.1, 2024. Special issue: ``Scientific Process Ontology and Metaphysics''. Url: \href{https://www.scielo.br/j/man/i/2024.v47n1/}{www.scielo.br/j/man/i/2024.v47n1}}

\author{Raoni Arroyo\thanks{Corresponding author. E-mail: \texttt{raoniarroyo@gmail.com}}}
\affil{Dipartimento di Filosofia, Comunicazione e Spettacolo, Università degli Studi Roma Tre, Rome, Italy.}
\affil{Centro de Lógica, Epistemologia e História da Ciência (CLE), Universidade Estadual de Campinas (UNICAMP), Campinas, Brazil}
\affil{Research Group in Logic and Foundations of Science (CNPq), Florianópolis, Brazil}

\author{Lauro de Matos Nunes Filho}
\affil{Universidade Federal de Santa Catarina, Departamento de Filosofia, Florianópolis, Brazil}
\affil{Research Group in Logic and Foundations of Science (CNPq), Florianópolis, Brazil}

\author{Frederik Moreira dos Santos}
\affil{Centro de Ciência e Tecnologia em Energia e Sustentabilidade (CETENS), Universidade Federal do Recôncavo da Bahia (UFRB), Feira dos Santos, Brazil}

\begin{document}
\sloppy\raggedbottom
\maketitle

\begin{abstract}
According to a particular interpretation of quantum mechanics, the causal role of human consciousness in the measuring process is called upon to solve a foundational problem called the ``measurement problem''. Traditionally, this interpretation is tied up with the metaphysics of substance dualism. As such, this interpretation of quantum mechanics inherits the dualist's mind-body problem. Our working hypothesis is that a process-based approach to the consciousness causes collapse interpretation (CCCI) ---leaning on Whitehead's solution to the mind-body problem--- offers a better metaphysical understanding of consciousness and its role in interpreting quantum mechanics. This article is the kickoff for such a research program in the metaphysics of science.
\paragraph{Keywords:} Consciousness causes collapse interpretation;
Metaphysics of science;
Philosophy of quantum mechanics;
Whiteheadian process metaphysics
\end{abstract}
\onehalfspacing

\section{Introduction}\label{sec:1}
\epigraph{``\textelp{} perhaps a theory of consciousness might shed light on the problems of quantum mechanics.''}{\citeauthor{chalmers1995consciousmind},``\citetitle{chalmers1995consciousmind}''.}

The empirical success of quantum mechanics is out of the question. Its conceptual success, on the other hand, is an entirely different story. Due to the measurement problem, quantum mechanics is conceptually incomplete; to conceptually complete quantum mechanics is to offer so-called ``interpretations of quantum mechanics''.\footnote{~But \textit{cf.} with \citet{arroyo-olegario2021,muller2015} for criticism of such terminology.} Interpretations of quantum mechanics, however, are strongly marked by \textit{ad hoc} hypotheses in the sense proposed by \citet[p.~986]{popper1974replies}, that is, ``a conjecture [is] `ad hoc' if it is introduced \textelp{} to explain a particular difficulty, but if \textelp{} it cannot be tested independently''.

Our focus here is on a particular solution to the measurement problem, namely: the \textit{consciousness causes collapse interpretation} \citep[hereafter, ``CCCI'' for short; see also][]{BarOas2016Consc}, which calls for the causal agency of human consciousness as a fundamental feature in measurement processes. Following the traditional Carnapian--Quinean metaontology in analytic philosophy, one might say that this interpretation is a framework that is ontologically committed to the existence of ``consciousness'' \citep{berto2015ontology,arenhartarroyo2021manu}. It is taken for granted, however, that this entity should be metaphysically understood under the umbrella of substance metaphysics and several strains of dualism \citep[see][]{arroyo-arenhart-2019}.

The problem is that dualism does not come for free: it inherits the mind-body problem which has haunted philosophers since Descartes. So it is safe to say that the CCCI \textit{also} inherits such a problem, so it becomes at the same time (a) puzzling why to adopt such an interpretation in the first place and (b) understandable why it is such an unpopular view among the quantum foundations' community. In \citeyear{poll2013schlo}, \citet*{poll2013schlo} presented a survey to the participants of a conference in quantum foundations containing multiple-choice questions on several open topics. There was a question about the role of the observer in physics, and only \textcolor{black}{6\% out of 33 participants (\textit{i.e.}, 2)} stated their belief that consciousness plays a crucial role in the measurement process. The results obtained by the survey, although not very expressive given the number of participants, are quite symbolic in terms of the attitude towards the concept of consciousness in quantum foundations.

This article aims to question the ---traditional--- direct link between the CCCI and the metaphysics of substance and dualism. By proposing a process-based approach to the CCCI, we aim to eliminate the crucial philosophical puzzles that come with it, \textit{viz.} the mind-body problem. Notably, the work of \citet{whitehead1928process} seems to be a good place to start, as there are already several attempts to understand quantum mechanics under the Whiteheadian, process-based, metaphysics \citep{malin2001naturelovestohide,shimony1964QMwhitehead,seibt2002quanta}. None of such attempts focused explicitly on the CCCI, however. And, as Whiteheadian metaphysics offers a non-eliminativist solution to the mind-body problem \citep{weekes2012mindbody}, we thought we should give it a try and connect both.

The article is structured straightforwardly. Section \ref{sec:old} presents (i) the measurement problem, (ii) the CCCI and its accompanying ontology, (iii) the traditional link with substance dualism, and (iv) its problems with the mind-body connection. Section \ref{sec:new} starts building up the process-based approach to the CCCI, focusing on Whitehead's process metaphysics. This particular section should be understood as a proposal/guideline for future research. Section \ref{sec:subs} discusses some metametaphysical advantages of process-based metaphysics over substance-based ones, applying our framework to Seibt's work in process metaphysics. 
As a disclaimer, we should add that, even though neither \textcolor{black}{Whitehead} nor Seibt discusses the CCCI in their work, we'll apply their metaphysical considerations to such an interpretation of quantum mechanics with the hope that their approach is general enough to encompass such a case. Section \ref{sec:conc} wraps it all up.

\section{Old direction: consciousness as substance}\label{sec:old}

Quantum foundations is the field in the philosophy of physics that deals with foundational questions in quantum mechanics, and it's hardly an exaggeration to state that it is almost entirely devoted to making sense of the measurement problem. In the foregoing, we'll proceed informally in presenting the problem and what amounts to its solution. It goes like this. Whichever way non-relativistic quantum mechanics is formulated, the equations of quantum dynamics obey a mathematical property called ``linearity''. This means, among other things, that: if we are to describe the state of a quantum system $Q_\psi$ in terms of its (say) position, and if the quantum system \textit{could be} located in regions $Q_A$ and $Q_B$, the state of $Q_\psi$ can be written as $Q_\psi=\alpha (Q_A)+\beta (Q_B)$, \textcolor{black}{with $|\alpha|^2$ and $|\beta|^2$ representing the probability for each result to occur.} This sum is called ``superposition'', and linearity implies this \textit{viz.} whenever two states are possible, you may add them, and this sum (\textit{i.e.} this superposition) is another possible state. \textcolor{black}{This is remarkably awkward from the conceptual point of view if one wants to assert whether the quantum system $Q_\psi$ has the property $Q_A$ (\textit{i.e.}, of being located at the region $A$) or not.}

One may attach a measuring device $M_\psi$ to check the state of $Q_\psi$; it turns out $M_\psi$ can measure (\textit{e.g.} point its pointer) to regions $M_A$ and $M_B$, each one corresponding to the positions of $Q_\psi$. Here's what linearity \textit{also} implies. Once $M_A$ and $M_B$ are \textit{possible} measurement outcomes of pointer positions, the state of the whole composite system (the quantum system and the measuring apparatus) should be described as a superposition of states, \textit{viz.}: \begin{equation}M_\psi Q_\psi=\alpha(M_AQ_A)+\beta(M_BQ_B).\label{eq:sup}\end{equation} This is to say that superpositions \textcolor{black}{don't} vanish by adding measuring devices.\footnote{\textcolor{black}{As an anonymous reviewer pointed out, this is not the general case, as a quantum-mechanical description may \textit{always} be rewritten in terms of a superposition of states in different bases. We are dealing with a specific kind of superposition, however, \textit{viz.}, macroscopically distinguishable superposition. In cases like these, the measurement result cannot be described in terms of a superposition \textit{simpliciter} ---a qualification of \textit{some kind} is required.}} Of course, not everyone espouses this position \citep[\textit{cf.} with][, for example]{bohr1928quantum}, but \textit{if} we are to apply the quantum dynamics for every physical system, this picture is inevitable \citep{vNeum1955mathematical}. If quantum mechanics is the only game in town, macroscopic measuring devices should obey its rules, just as microscopic quantum systems. Adding further measuring devices to measure measuring devices (sic) won't give a way out of further superposed states, and adding human observers will similarly add one more term to the superposition. This is called von Neumann's chain \citep{desp1999concep}. On the standard way of thinking, there's no physical representation of states such as the one described in Equation \ref{eq:sup}. Within this standard way of thinking, as \citet{albert1992quantum} emphasizes, the superposed state is:
\begin{quote}
    \textelp{} \textit{a state in which there is no matter of fact about whether or not} [the observer] \textit{thinks the pointer is pointing in any particular direction.} \citep[p.~79, original emphasis]{albert1992quantum}.
\end{quote}

But this is abstruse. Such a description contradicts our immediate experience of the world. We experience pointers pointing to a determinate pointer position, either $M_A$ or $M_B$, \textit{exclusively}; and if the measuring device is well-calibrated ---if it's a \textit{good} measuring device--- it will point to the region of space corresponding to the actual position of the quantum system: again, \textit{exclusively} either $Q_A$ or $Q_B$. End of story. But the linearity property of quantum dynamical equations prevents just that. This is the measurement problem.

Thanks to \citet{Maudlin1995measurementproblem},\footnote{~But \textit{cf.} with \citet{muller2023}.} the now-standard way of classifying the solutions to the measurement problem is like this. As per the abstruse description of physical states given by Equation \ref{eq:sup} is not on par with our experience, we must admit at least one thing out of three. Either: (i) that something is missing in this description, \textit{i.e.}, this equation is not giving us enough information about the physical systems of our interest; (ii) that the trouble is with the \textit{linearity}, and that there is another dynamics besides the linear one, \textit{viz.}, the collapse; (iii) that every term described by Equation \ref{eq:sup} happens to be the case \textit{somewhere else}, \textit{e.g.}, in another world.

The CCCI solves this problem by pursuing the second strategy.\footnote{~For a philosophical survey on quantum interpretations, the reader should be referred to \citet{Jammer1974,albert1992quantum}. Here, we'll discuss just one interpretation of quantum mechanics.} It puts the minds of quantum observers outside the scope of linear quantum dynamics, \textit{i.e.} outside the scope of superpositions. The collapse, \textit{e.g.}, the transition from $M_\psi Q_\psi=\alpha(M_AQ_A)+\beta(M_BQ_B)$ to ---exclusively--- $M_\psi Q_\psi=(M_AQ_A)$ or $M_\psi Q_\psi=(M_BQ_B)$ (up to a probability) is caused by the interaction with a human mind \citep{wigner1961mindbody}.

\color{black}
The motivations for such a proposal are nicely listed by \citet{chalmersmcqueen2022}:

\begin{quote}
    \textelp{} the [CCCI] view provides one of the few interpretations of quantum mechanics that takes the standard measurement-collapse principle at face value. Other criteria for measurement may be possible, but understanding measurement in terms of consciousness has a number of motivations. First, it provides one of the few non-arbitrary criteria for when measurement occurs. Second, it is arguable that our core pretheoretical concept of measurement is that of measurement by a conscious observer. Third, the consciousness collapse [CCCI] view is especially well-suited to save the central epistemological datum that ordinary conscious observations have definite results. Fourth, understanding measurement as consciousness provides a potential solution to the consciousness-causation problem: consciousness causes collapse. \citep[p.~12--13]{chalmersmcqueen2022}.
\end{quote}

Yet, recall that this is an unpopular proposal \citep*{poll2013schlo}. There are many reasons for this. \citet[p.~13, original emphasis]{chalmersmcqueen2022} point out that the CCCI's unpopularity among the scientific community might have to do with its popularity among the \textit{un}scientific literature and its link with Eastern religious traditions; ``more substantively,'' they continue, ``the view is frequently set aside in the literature on the basis of \textit{imprecision} and on the basis of \textit{dualism}.''

Let's briefly analyze each of such charges in turn, beginning with the charge of imprecision, as it was eloquently put by \citet{bell2004-1981}, \citet{albert1992quantum} \citet{becker2018} respectively:

\begin{quote}
    It would seem that the theory is exclusively concerned with `results of measurement' and has nothing to say about anything else. When the `system' in question is the whole world where is the `measurer' to be found? Inside, rather than outside, presumably. What exactly qualifies some subsystems to play this role? Was the world wave function waiting to jump [collapse] for thousands of millions of years until a single-celled living creature appeared? Or did it have to wait a little longer for some more highly qualified measurer ---with a Ph.D.? \citep[p.~117]{bell2004-1981}.
\end{quote}

\begin{quote}
    How the physical state of a certain system evolves (on this proposal) depends on whether or not that system is conscious; and so in order to know precisely how things physically behave, we need to know precisely what is conscious and what isn't. What this ``theory'' predicts will hinge on the precise meaning of the word conscious; and that word simply doesn't have any absolutely precise meaning in ordinary language; and Wigner didn't make any attempt to make up a meaning for it; so all this doesn't end up amounting to a genuine physical theory either. \citep[p.~82--83]{albert1992quantum}.
\end{quote}

\begin{quote}
    Stating that consciousness collapses wave functions does arguably solve the measurement problem but only at the price of introducing new problems. How could consciousness cause wave function collapse? Since wave function collapse violates the Schrödinger equation, does that mean that consciousness has the ability to temporarily suspend or alter the laws of nature? How could this be true? And what is consciousness anyhow? Who has it? Can a chimp collapse a wave function? How about a dog? A flea? ``Solving'' the measurement problem by opening the Pandora's box of paradoxes associated with consciousness is a desperate move, albeit one that seemed reasonable at the time, in the absence of other fully developed solutions to the measurement problem. \citep[p.~75]{becker2018}.
\end{quote}

Point blank, all these famous quotations call attention to the fact that it is simply implausible that the state of any given physical system remains undetermined (\textit{i.e.}, without any determinate fact of the matter) and devoid of physical meaning ---in a state of `suspension', as it were--- until a `conscious' system observe it.

In fact there's an increasingly heated debate in the metaphysics of science concerning such matters, \textit{viz.} how to interpret the absence of determinate properties in specific situations of quantum superposition; the traditional name for such a feature/problem is ``quantum metaphysical indeterminacy''. See \citet{torza2023} for a recent survey and updated reference list. In particular, such indeterminacy in \textit{e.g.} location properties was recently interpreted by \citet{glick2017} as an existential indeterminacy in the so-called ``sparse view'' on quantum metaphysical indeterminacy called \citep[see the debate in][]{calosi-wilson2021phil,torza2022,glick2022}. And each interpretation of quantum mechanics responds to such indeterminacy in its own way ---sometimes getting rid of indeterminacy altogether \citep{calosimariani2021}. The implausibility of the CCCI has nothing to do with its philosophical unsoundness. Every quantum interpretation might be unsound to the eyes of the working physicist. 

Yet, the measurement problem must be addressed somehow ---if only we want to take ontological lessons from quantum mechanics. \textit{If} we want to consistently interpret what it might be telling us about what the world could be ---that is, \textit{if} we want to know, as per \citet[p.~3433]{ruetsche2015shaky} ``an account of what the worlds possible according to [quantum mechanics] are like''--- \textit{then} we must solve the measurement problem. Which is to give a precise account of how we perceive determinate, non-overlapping states as measurement outcomes. And the introduction of a causal conscious agency is one such solution. Other solutions are available as well, and nothing forces one upon the CCCI. Maybe each term of the superposition inhabits a different branch of the universal wave function, or a different Everettian world in the Everettian multiverse ---who knows? But yet, scientific realism is not our matter here.

So let us pass to the charge of \textit{dualism}. At first sight, to say that a view is `dualist' is hardly a criticism \citep[see][for a similar argument]{arroyo-arenhart-2019}. Self-avowed dualists certainly won't feel the pull of such a criticism. But then we might reconsider that. Because dualism, however, has a weak spot. Namely, the mind-body problem. So maybe the charge of dualism might do the damage that \citet{chalmersmcqueen2022} warned us a few paragraphs ago. After all, swapping the measurement problem with the mind-body problem is not a safe bet.
\color{black}
So here's our working hypothesis: perhaps the implausibility of the CCCI is closely related to the absence of its metaphysical development. Thus, formulating metaphysical theories tailored-made for quantum mechanics seems to be an essential task for contemporary philosophy. Indeed, there's even a methodology for doing so: the Toolbox Approach to metaphysics (see \cite{french2012toolbox,FrenMcken2015toolboxagain}; \cite{french2014structure,french2018toolbox}, and \textit{cf.} with \cite{arenhartarroyo2021veritas}) suggest that philosophers of science should use the theoretical devices produced by analytic metaphysics as a source to obtain a better understanding of scientific theories.\footnote{~This is what, \textit{e.g.}, \citet{wilson2020QMmodal} did by presenting a Lewisian-based quantum-mechanical version of modal realism ---called ``quantum modal realism''---, by taking into account (another) specific interpretation of quantum mechanics at face value and developing in detail a tailored-made metaphysical view for it (see \cite{arroyo2021}, but \textit{cf.} with \cite{arroyoarenhart2022foop}).}

If this diagnostic is sound, here's \textit{another} working hypothesis that we'll run throughout this article: the hypothesis that the elaboration of a metaphysics for the notion of ``consciousness'', inspired by the metaphysics of processes as presented by Alfred North \citeauthor{whitehead1928process} in his \textit{magnum opus} ``\citetitle{whitehead1928process}'' \citeyearpar{whitehead1928process}, could shed new light to the CCCI. Thus, it is a proposal based on the hope that, as \citet[p.~311]{chalmers1995consciousmind} points out, ``even if quantum mechanics does not explain consciousness, perhaps a theory of consciousness might shed light on the problems of quantum mechanics''.

As we're not the first to think about this, let us see where others left off so we can pick it up from there. \citet[p.~271]{shimonymalin2006dialogue} ponder different attitudes towards the interpretation of the concept of measurement and consider that the CCCI \textit{could be} favorable for a \textcolor{black}{Whiteheadian} philosophy. In fact, \citet{shimony-wigner1961letter} tried to convince Wigner that his interpretation was not necessarily tied up with substance dualism, and that could be fruitfully interpreted on a Whiteheadian framework. However, this was just hinted by these authors, as they end up denying the plausibility of this interpretation due to its ---apparent--- commitment to the idea of \textit{subjective consciousness}. \citet[p.~763--767]{Shi1963role}, like most physicists today, rejected interpretations that consider the subjective consciousness of the observer to be the causal agent of the collapse in quantum measurement. We'll pick up from where \citet{shimonymalin2006dialogue} left. Contrary to them, we will bite some bullets and try to develop the Whiteheadian version of the CCCI a little further.

Bluntly put, while the CCCI has been dismissed on many grounds,\footnote{~See, for instance, \citet{albert1992quantum,lewis2016quaont,chalmers1995consciousmind}.} it is not so quickly ruled out and remains a viable interpretation of quantum mechanics \citep[see][]{arroyo-arenhart-2019}. Nevertheless, it inherits a severe philosophical problem \textit{viz.} the mind-body problem. Perhaps the most explicit statement of the traditional connection between the CCCI and the mind-body problem comes from \citet*{shimony-etal1977}:

\begin{quote}
\textelp{} if [the measurement problem] does pose a genuine problem, then it is a very hard one, and some physicists and philosophers have come to believe that no easy, nonradical solution will succeed. Since the mind-body problem is a perennial unsolved problem (which classical physics somehow managed to bypass without solving), one could conjecture that the two problems are intermeshed. \citep*[p.~761]{shimony-etal1977}.
\end{quote}

So here's how things stand to the CCCI: even if we have no physical nor metametaphysical grounds to objectively rule it out \citep[again, see the whole argument in][]{BarOas2016Consc,arroyo-arenhart-2019}, and even if one bite all the bullets coming from the philosophical implications of such an interpretation, one thing indeed stands: \textit{if} the CCCI is tied up with substance dualism, \textit{then} it is stuck with the mind-body problem. That is, once this interpretation is adopted, it traditionally inherits the burden of proof of solving the mind-body problem, \textit{viz.}, to give a precise account of how a non-physical mind can interact with a physical system \textit{e.g.}, a measuring apparatus, or any other quantum-mechanical system. By doing so, however, the CCCI exchanges the measurement problem in quantum mechanics with the mind-body problem in philosophy. The odds are against such an interpretation, as the latter last unsolved from \textit{much more time} than the former.

Well, perhaps.

\section{New directions: consciousness as process}\label{sec:new}

Here's another way to state the problem. The literature espousing the CCCI of quantum mechanics presupposes, directly or indirectly, a dualist metaphysics for the concept of consciousness that is, at the same time, (i) dualist, insofar as it separates consciousness and ``matter'' into distinct substances, and (ii) subjectivist, to the extent that the notion of consciousness is based on the ``I'', which thinks and therefore exists.

Now, here's the possible way out. Unlike materialist metaphysics, Whiteheadian metaphysics is considered non-reductionist as it does not deny the causal efficacy between the material and non-material (mental) poles of existence. Unlike dualists nor does it consider them ontologically separate. In Whitehead's metaphysical model, consciousness contains and is contained by the concept of matter; \textcolor{black}{from a perspective of processes (and not objects), consciousness transcends and is transcended by matter \citep[see also the quotation below from][p.~175]{griffin2009whiteheadconsc}.} Thus, it can be stated that, from a process metaphysics perspective, the world is both immanent and transcendent. At first, such categorizations eliminate the main difficulties the concept of consciousness faces. However, the aspect of subjectivism considered above (ii) needs to be taken into account since a subjectivist interpretation is undesirable in a scientific theory, and Whitehead considers that the concept of consciousness has a subjective aspect ---it is not, however, \textit{reduced} to subjectivity as in dualistic metaphysics \citep[see][]{griffin2001process}. Bearing in mind that Whitehead's model offers an original way ---and little mentioned in the specific literature, as \citet{weberweekes2009neglected} point out--- of dealing with the problem mentioned above, we argue that Whiteheadian-like metaphysics could be fruitful to the notion of consciousness as applied to the interpretation of quantum mechanics. This section introduces a framework for such a development in the metaphysics of science.

While using Whitehead's process metaphysics to interpret the relation between consciousness and quantum mechanics is new \citep[see][]{gao2022}, the broader attempt to interpret quantum mechanics from certain aspects of Whitehead's philosophy is not new. In fact, the results of physics were one of the main starting points for \citet[p.~121--122]{whitehead1928process}'s theory, which aimed to provide a conceptual basis for what it refers to as ``quantum theory''. However, as \citet[p.~240]{shimony1964QMwhitehead} noted, the aforementioned ``quantum theory'' in Whitehead's is the early quantum theory, \textit{viz.} the theory as first developed in the early 1900s. The period in which Whiteheadian philosophy was being developed preceded a period of significant changes in quantum mechanics, including debates about the foundations and the ontology associated with their interpretations ---especially in the 1930s. Thus, it is very unlikely that Whitehead mentioned in his writings the most ``recent'' developments in quantum mechanics, relative to its contemporaneity. Taking this into account, it is natural that authors such as \citet{shimony1964QMwhitehead} and \citet{malin1988whiteheadbell} propose some modifications in the concepts of Whiteheadian metaphysics to accommodate the interpretation of quantum mechanics.

Perhaps the first documented proposal to use Whiteheadian philosophy to elucidate the debate around interpretations of a relatively better established quantum theory was that of \citet{burgers1963qm, burgers1965qm}, followed mainly by \citet{Shi1963role, shimony1964QMwhitehead,stapp1979bell, stapp1982mind,malin1988whiteheadbell, malin1993collapse, malin2001naturelovestohide,epperson2004QMwhitehead,ferrari2021,seibt2002quanta}. It's worth emphasizing that all the referred authors use the same concepts to make the parallel between quantum mechanics and the metaphysics of \citet{whitehead1928process}:

\paragraph{1)} Regarding quantum mechanics, we highlight the concept of ``\textit{potentia}'' contained in the later writings of \textcolor{black}{\citet[p.~41]{heisen1958physphil}}, who interprets the concept of ``quantum state'' as a tendency, something between the idea of the phenomenon (or event) and its actuality, a ``kind of physical reality just in the middle between possibility and reality''. Although \textcolor{black}{\citet[p.~51]{heisen1958physphil}} elaborates his concept of ``\textit{potentia}'' as a reinterpretation of the Aristotelian concept of \textcolor{black}{``potentiality'' \citep[``\textit{dunamis}'' in Greek, see][\S~12 for overview and references with regard to Aristotelian Metaphysics]{sep-aristotle-metaphysics}}, \citet[p.~263]{shimonymalin2006dialogue} guarantee that such a proposal is original since no other metaphysics until then would have proposed this modality for reality. In \citeauthor{heisen1958physphil}'s \textcolor{black}{\citeyearpar[p.~158]{heisen1958physphil}} conception, even contrary potentialities could coexist, as in the case of superposition, ``since one potentiality may involve or overlap with other potentialities''. As \citet[p.~264]{shimonymalin2006dialogue} point out, the very concept of ``superposition'' would be ``derivative from the fundamental metaphysical innovation of potentiality''.

In such an interpretation, a measurement consists in the actualization, by means of the collapse, of one among many superposed possibilities.
\begin{quote}\textelp{} the theoretical interpretation of an experiment requires three distinct steps: (1) the translation of the initial experimental situation into a probability function; (2) the following up of this function in the course of time; (3) the statement of a new measurement to be made of the system, the result of which can then be calculated from the probability function. \textelp{} It is only in the third step that we change over again from the ``possible'' to the ``actual''. \citep[p.~46--47]{heisen1958physphil}.\end{quote}
This certainly diminishes the idealistic tone of \citet[p.~73]{heisenberg1927uncert} when he states that an event ``\textelp{} comes into being only when we observe it''. This Heisenbergian feature was termed by \citet{sep-qt-uncertainty} as ``measurement=creation''. In the Whiteheadian context, we find it more appropriate to call it ``\textit{measurement=actualization}''. \citet[p.~76--77]{malin2003col} points out that the potentialities are not to be understood as events in space-time ---this would be a characteristic of \textit{actualities}.

\paragraph{2)} Regarding Whitehedian metaphysics, the concept of ``actual entities'' is central to the interpretation of quantum mechanics. Whitehead enunciates such a concept for the first time as follows: \begin{quote} ``Actual entities'' ---also termed ``actual occasions''--- are the final real things of which the world is made up. There is no going behind actual entities to find something more real. \citep[p.~18]{whitehead1928process}. \end{quote} According to Malin, the concept of ``actual entities'' would be the base of the metaphysics proposed by Whitehead. Given the scope and objectives of this text, it is impossible to summarize all of Whitehead's philosophical construction. We follow the outline proposed by \citet[p.~77--78]{malin1993collapse};\footnote{~See also \citet[p.~266--267]{shimonymalin2006dialogue}.} which highlights eight central aspects, relevant to the debate over the interpretation of quantum mechanics; out of the eight aspects, we select only four that we consider specifically relevant to the concept of measurement:

\begin{enumerate}
    \item An actual entity is a process of timeless and creative ``self-creation'', which leads to a momentary appearance of the actual entities in space-time;
    \item Actual entities are instantaneous; after the single instant in which they emerge in space-time through self-creation, they merge again (in Whitehadian terminology, they ``prehend'') into a timeless and out-of-space domain with all current entities (past and future), as potentialities;
    \item Every actual entity is related to and interconnected (in Whiteheadian terminology, forms a ``\textit{nexus}'') with all actual entities;
    \item The end of the self-creation process of an actual entity, \textit{i.e.}, its momentary appearance in space-time, is the self-creation of a new actual entity or an ``pulse of experience'', so that the Whiteheadian universe is not an actual universe of ``objects'', but a universe of ``experiences''.
\end{enumerate}

As \citet[p.~92]{stapp2007whiteheadQM} points out, the parallel between the metaphysics of \citet[p.~72]{whitehead1928process} in which ``the actual entities \textelp{} make real what was antecedently merely potential'' and Heisenberg, in which ``\textelp{} the transition from the `possible' to the `actual' takes place during the act of observation'' is very suggestive. For Shimony, such a parallel could be visualized as follows:

\begin{quote}
Consider, for simplicity, two entangled particles. If they are regarded, together, as a single actual entity, their mutual dependence is natural: both arise out of a single field of potentiality. When a measurement takes place on either particle, it breaks the connection, creating a relationship between two actual entities \textelp{}. \citep[p.~274]{shimonymalin2006dialogue}.
\end{quote}

For \citet[p.~81]{malin2003col}, the gain of such an interpretation is to offer a new horizon of answers to the following question ---not yet answered--- in the debate about the interpretation of quantum measurement: \textit {what is the mechanism of collapse?}. In Whiteheadian metaphysics, the universe would not be a universe of objects (or fields), but a universe of experiences or processes, so that if the collapse axiom is interpreted as the process of self-creation of an actual entity, such a process could not be a mechanism that excludes the possibility of creativity. In this reading, the concept of ``mechanism'' seems to have no place. This is the central point we want to emphasize in this particular reading of consciousness and its role in quantum measurement: if there is no need for a mechanism, there's also no need to search for the external cause of the collapse.  Consciousness, as an occasion of experience, may self-create actualities. Regarding the interpretation of causal consciousness, \citet[p.~260--261]{malin2001naturelovestohide} rejects the interpretation that consciousness plays a causal role in the collapse.

\color{black}
With all the previous conceptual discussion under our belts, let's try to cook up an example of how things might turn out in this framework. In it, both terms described by a superposition (\textit{e.g.}, equation \ref{eq:sup} reproduced below) are potentialities outside space and time. They're in the Aristotelian realm of \textit{potentia}, and quantum-mechanical descriptions capture the evolution of such overlapping potentialities. \begin{equation*}M_\psi Q_\psi=\alpha(M_AQ_A)+\beta(M_BQ_B).\end{equation*} If $M_AQ_A$ and $M_BQ_B$ are complex enough to be understood as potentialities representing macroscopically distinguishable states, then \textit{a pulse of experience} \citep[see][]{nobo2003whiteheadQM} might tell them apart\footnote{~\textcolor{black}{Not necessarily \textit{human} experience; more on that below.}} and hence an actual entity (or actual \textit{event}) might be conceived ---and the probabilities $|\alpha|^2$ and $|\beta|^2$ weights them. Such an actualization of potential entities into actual entities is the collapse, which is an event outside space-time. All in all, with this we've tried to push a little further Abner Shimony's idea according to which:

\begin{quote}
    \textelp{} quantum mechanics points to the need for profound changes in our understanding of space-time structure. Nothing so simple as just discretizing space-time structure is envisioned here. No, for Abner [Shimony] the question has always been the Whiteheadean or Aristotelian one about the actualization of potentialities. \citep[p.~8]{howard2009}.
\end{quote}

Notice that there is an element of vagueness in such a framework, which is similar to the Everettian proposal when we think of the `threshold' of such a transition from potentiality to actuality:

\begin{quote}
    When we say that a world ``splits'' or ``branches'' (for instance, in the course of a measurement experiment), we are actually talking about a gradual process. Think of a wave packet on an extremely high-dimensional configuration space fanning out into two or more parts that become more and more separated in that space. Don't try to think of an exact moment in which it goes ``bing'' and the world suddenly multiplies. The concept of a ``world'' has a certain vagueness ---it's not possible, in general, to say exactly how many worlds exist or at what moment in time a new splitting has occurred. \citep[p.~118]{durrlazarovici2020}.
\end{quote}

The process of collapse (\textit{qua} the creation of actual entities) is then similar to the process of branching in the Everettian solution to the measurement problem. The gradual increasing complexity is analogous to the gradual separation of states in Everettian quantum mechanics. But there's also an analogy with the spontaneous collapse model from \citet*{grw}. In the GRW model, new constants of nature are introduced to solve the measurement problem. One of them is the ``collapse rate'' which increases with the amount of elements in the scope of the quantum system. This would explain why we find collapsed states when we attach a measurement apparatus (containing a large number of quantum components, hence a very complex system), and why we find interference patterns of simpler systems (\textit{viz.}, containing few quantum components) such as individual particles quantum entities in a two-slit setup. As should be expected, the Whiteheadian version of the CCCI ---just like the GRW model--- also holds that `experience' emerges in complex/decoherent systems. Future studies in the field of neuroscience might shed light on the degrees of experience, specifically the ``Integrated Information Theory'' of consciousness \citep[``IIT'', see][]{tononi2004,tononikosch2015,tononi-etal2016} ---just as it is applied in other contemporary versions of the CCCI, \textit{e.g.} \citet{chalmersmcqueen2022}--- as the IIT furnishes a framework capable of explaining how consciousness emerges from seemingly `unconscious' objects:

\begin{quote}
    IIT was not developed with panpsychism in mind (sic). However, in line with the central intuitions of panpsychism, IIT treats consciousness as an intrinsic, fundamental property of reality. IIT also implies that consciousness is graded, that it is likely widespread among animals, and that it can be found in small amounts even in certain simple systems. Unlike panpsychism, however, IIT clearly implies that not everything is conscious. \citep[p.~11]{tononikosch2015}.
\end{quote}

\color{black}
In particular, the above-mentioned increasing complexity that enables experience does not require human experience to enable the creation of an actual entity. This might beg the question of whether our framework entails panpsychism just as Whitehead's own metaphysics does \citep[see][]{skrbina2017}. As we understand it, the Whiteheadian version of CCCI might not \textit{require} panpsychism (although it is certainly \textit{compatible} with it, see \cite{chalmersmcqueen2022,okonsebastian2022}). If it is maintained that the \textit{experience} is the gradual process from non-space-time potentialities and space-time actualities (again, don't try to think it of a threshold, otherwise we'll stumble over the old problems), then we don't need to think of the CCCI meaning \textit{phenomenal} consciousness (\textit{i.e.} subjective/\textit{qualia} internal-experience, `what-is-it-like-to-be-an-electron' kind of consciousness), but \textit{access} consciousness would do the job (\textit{i.e.}, the ability of interaction with other states). And here we align with \citet{debarrosmontemayor2022}, hence \textit{panexperimentalism} would be a more appropriate label for this framework. By doing so, we disagree with \citet[p.~12]{chalmersmcqueen2022}, to whom  ``[b]y consciousness, what is meant is phenomenal consciousness, or subjective experience.''

Notably, the study of the notion of consciousness is permeated by polarities, bequeathing to contemporary discussion the same scope of theoretical options given centuries ago: either a form of reductionist monism (of which the theses of materialism and epiphenomenalism are the most popular) or dualism. For Shimony, a Whiteheadian-inspired metaphysics can offer a fruitful approach to the traditional mind-body problem:
\color{black}

\begin{quote}
There's nothing we know better than that we have conscious experience. There's nothing that we know much better than that the matter that the world is made of is inanimate. \textelp{} Put those together; you don't have a solution, you have a puzzle, a terrible puzzle. \textelp{} I'm very sympathetic with Whitehead because Whitehead does give an answer to this by postulating a primitive universe which is not entirely inanimate; he calls his philosophy the ``philosophy of organism.'' That is as promising as anything I know for a solution to the mind-body problem but it leaves out the details terribly. \citep[p.~451--452]{shimonysmolin2009dialogue}.
\end{quote}

Briefly, \citet{whitehead1928process} solves the mind-body problem by proposing a holistic theory of reality that recognizes the interdependence between physical and mental aspects of experience. By rejecting the dualist notion that mental and physical are \textit{distinct substances}, instead, it is argued that both mental--physical poles are part of a continuum of experience that encompasses all aspects of reality. According to this view, the fundamental units of reality are not substantial material particles but processes or \textit{occasions of experience} that include both physical and mental aspects. These occasions of experience are constantly interacting with each other, forming an ever-evolving web of interconnected experiences. Overall, Whitehead's theory of reality offers an integrated view of mind and body, avoiding dualism that has haunted philosophy for centuries. However, we should address the ``details'' to which Shimony refers in the above passage ---which are also mentioned by Malin in the form of still open problems within Whiteheadian metaphysics:

\begin{quote}
Whitehead's process philosophy provides a metaphysical foundation for the understanding of reality. It leaves, however, essential questions unanswered: Does reality consist of levels, some of which are ``higher'' than others in a profound sense? Do human beings have a place, and a role to play, in the cosmological scheme? \textelp{} surprisingly, the mysterious ``collapse of quantum states'' continues to be a rich source of suggestions. The collapse, the process of transition from the potential to the actual, involves a selection: There are many possibilities, only one of which is actualized. How is the selection made?
\citep[p.~189]{malin2001naturelovestohide}.
\end{quote}

The proposal put forth by \citet[p.~93]{malin2003col} would be to follow the maxim, attributed to Paul Dirac, that ``Nature makes the choice'', that is, that ``Nature'' causes the collapse. Although the definition of this ``Nature'' is not specified with a capital letter, in its reading, this corresponds to the actualization of potentialities, or even, its self-creation, with intrinsic randomness ---hence the quantum indeterminacy. Given the investigative character of this proposal, it seems premature to align ourselves with such a perspective beforehand. So let us see another candidate.

Another attempt to interpret quantum mechanics, in particular, the causal role of consciousness in quantum measurement, is made by Henry Stapp. His proposal goes in the opposite direction to that proposed by the interpretation of causal consciousness, which sought to use consciousness to understand quantum mechanics; \citet{stapp2007whiteheadQM} seeks to use quantum mechanics to understand consciousness ---a path that is also traced by \citet{penrose1994shadows}. However, as observed by \citet[p.~172]{landau1998errorpenrose}, ``Penrose accepts that the conscious mind arises as a functioning of the physical brain \textelp{}'', a thesis that is not endorsed by \citet{stapp2006quantumdualism}, who proposes a metaphysics he calls \textcolor{black}{``interactive dualism''}. As Mohrhoff points out:

\begin{quote}
The theory which he [Stapp] ends up formulating is completely different from the theory he initially professes to formulate, for in the beginning consciousness is responsible for state vector reductions [collapse], while in the end a new physical law is responsible ---a law that in no wise depends on the presence of consciousness. \citep[p.~250]{mohrhoff2002stapperror}.
\end{quote}

It is fair to say that \citet[p.~264]{stapp2002response} itself states that ``[this] is not my final theory''. Yet when asked by Malin whether Stapp's theory considers, as a consequence, that consciousness causes collapse, Stapp categorically replies that he does not endorse such an interpretation \citep[see the full dialogue in][p.~ 110]{eastman2003QMwhitehead}. It is possible to interpret the Whiteheadian ontology from dualistic metaphysics. According to \citet[p.~169]{lovejoy1960againstdualism}, Whitehead would be ``an opponent of the dualism we are concerned with here, but only a dualist with a difference''; as \citet{shimony1964QMwhitehead} points out, the dualist reading, if legitimate, would be fundamentally contrary to Whitehead's own proposal which, as emphasized by \citet{weekes2009caus}, is essentially monist.

Understanding the plurality of readings (dualists and monists) of Whiteheadian metaphysics, we tried to use the monist reading key, offered by \citet{weekes2012mindbody}, \citet{griffin2009whiteheadconsc} and \citet{nobo2003whiteheadQM} to understand the concept of consciousness \textcolor{black}{as related to} the notion of ``collapse'' in the interpretation of the concept of measurement in quantum mechanics. As \citet{griffin2009whiteheadconsc} points out, the Whiteheadian conception of consciousness differs radically from the dualist and physicalist (reductionist) positions ---which are the predominant readings for the concept of consciousness in the philosophy of physics--- even though it maintains some aspects of these metaphysical conceptions:

\begin{quote}
With dualists, Whitehead agrees that consciousness belongs to an entity ---a mind or \textit{psyche}---that is distinct from the brain. That genuine freedom can, partly for this reason, be attributed to conscious experience. With materialists, Whitehead shares a naturalistic sensibility, thereby eschewing any even implicitly supernaturalistic solution to philosophical problems, and, partly for this reason, rejects any dualism between two kinds of actualities. Like materialists, in other words, he affirms a pluralistic monism. He thereby regards consciousness as a function of something more fundamental. \citep[p.~175]{griffin2009whiteheadconsc}.
\end{quote}

\citet[p.~225]{nobo2003whiteheadQM} also emphasizes that the notion of consciousness, in Whiteheadian metaphysics, is not reduced to human experience or subjectivity. Furthermore, as observed by \citet[p.~206--208]{katzko2009processconsc}, the contemporary debate in the philosophy of mind, specifically regarding the reading of the notion of consciousness, is, in its most expressive part, committed to a materialist metaphysics or dualistic. By way of sampling: there are proponents of a physicalist metaphysics who, like \citet{stapp1982mind}, consider mental causation over the physical but, at the same time, consider brain structure as definitely crucial for the occurrence of the mental aspect; \citet{dennet1991consciousnessexpl}, even more radical, defends the thesis of ``functionalism'' that the mind is a product of the cerebral arrangement, not being able to have a causal action on the brain, placing himself between the materialists or epiphenomenalists; \citet{chalmers1995consciousmind} considers both poles, material and mental, equally important, which brings him closer to the dualists through what he calls ``interactive dualism''; in all cases, one of the central questions would be one of causation, that is: \textit{how could the physical aspect of reality give rise to the mental aspect?}

These dualist difficulties are undone in the monist, non-reductionist metaphysics offered by Whitehead \citep[see, again,][]{weekes2012mindbody}; as a corollary, a Whiteheadian, process-based metaphysics for consciousness in quantum mechanics will not suffer the same fate as its dualist cousins. 

In the following, we'll argue that one may trace useful parallels between our proposal and contemporary process metaphysics as per the work of Johanna Seibt.

\section{Processes ontology and measurement: beyond dualistic substantialism}\label{sec:subs}

Many doors are open by Whitehead's approach toward consciousness; however, many windows still stay closed. The ``details'' missing from his theory may not result from insufficient conceptual attention but from adequate ways to approach the matter. Only recently, the theory of processes regained a new boost: this controversial ontological category has reentered the room of ontology in the last decades, and today one of its prominent defenders is \citet{seibt2002quanta,seibt1990towards,seibt2004free}. This way, we seek to renovate the efforts to understand the relation between consciousness and collapse from the point of view of analytical ontology. In doing so, perhaps we may try to fulfill some missing details in Whitehead's metaphysics.

\citet[p.~483]{seibt1990towards} proposes a non-Whiteheadian view on processes, by suggesting a process ontology that seeks to escape from the picture according to which there exist some \textit{thisness} underlying entities in a given domain. Such a picture is grounded on the assumption that all entities can be reduced to well-delimited entities at some space and time or depend on something to exist (\textit{i.e.}, the substance picture). \citeauthor{seibt1990towards} \citeyearpar[p.~500]{seibt1990towards}Seibt's approach goes in the opposite direction, arguing that it is possible to make ontology without being spelled by the myth of substance. Additionally, she upholds the view that processes might be applied to investigating entities based on mathematical characterizations, which is the case of quantum entities \citep[p.~87]{seibt2002quanta}. Indeed, her approach toward quantum theories is restricted to the characterization of how free processes could model an observable.

Bluntly put, \citet[p.~87]{seibt2002quanta} has a two-folded procedure toward ontology: first, she considers that ontology must be built in an agentive way, which means that ontology relies on performative aspects of a group or activity; second, she conceives ontology as a reductive ontology of processes, which is ``\textelp{} well-founded, formally simple, and monocategorial''. According to \citet[p.~483]{Johanna-2009}, the category type might be defined by different features \textit{e.g.}, particular, individual, complex, fuzzy, \textit{etc.}

Seibt's work conserves some similitude with the Whiteheadian project. Free processes are applied to modes of occurrence in space and time. Besides, her agentive approach toward ontology and language opens the same path aimed by Whitehead (\textit{viz.}, intersubjectivity). Both pertain to different thought paradigms; still, they are relatable at different levels. For example, Seibt's crusade against the concept of substance boosted her ontology towards an ontology decentered from substantialism and its immediate sibling particularism.

Free processes are subjectless activities, and activities are concrete non-countable individuals \citep[p.~83-84]{seibt2002quanta}. After all, if free processes were subjects, they could be predicated and automatically be particularized as substances. Above all, ``\textelp{} free process are \textit{not particulars}'' \citep[p.~85]{seibt2002quanta}, but they still are concrete individuals:
\begin{quote}
Free processes are (i) concrete or spatio-temporally occurrent (ii) individuals that are (iii) `dynamic stuffs' rather than changes in a subject, (iv) They are non-particulars or (contingently) multiply occurrent. (v) They are not fully determinate, \textit{i.e.}, they have different degrees of specificity or determinateness. (vi) Simple free processes are not directed developments (events) but are dynamically homeomerous. \citep[p.~86]{seibt2002quanta}.
\end{quote}

Since free processes cannot be in a subject-property relation, \citet{seibt2002quanta} defines \textit{homeomerity} as a mereological relation in which each part of an entity is structurally the same as the whole entity.  Free processes are homeomerous; that is to say, all their parts are still free processes. 

Having briefly introduced the notion of ``free process'', let's get back to the heart of the matter, \textit{viz.} measurement in quantum mechanics. The measurement process of something occurs in analogy with a classical measurement. Such a proposal efficiently apprehends the indeterminacy of the outcomes of a measurement process. Here she introduces the disposition category to describe the process before the measurement. Such a category might be related to Heisenberg's notion of \textit{potentia}; both are used similarly. After all, in tone with Heisenberg and Whitehead, \citet[p.~87]{seibt2002quanta} understands ``potentialities'' as one of many ``\textelp{} species of dispositions \textelp{}'' such as ``capabilities, capacities, tendencies, or propensities.''.

At this point, \citet{seibt2002quanta} provides us with a well-founded ontological approach toward measurement and quantum entities. However, it is still open whether this approach can be extended to the CCCI by connecting consciousness and collapse. Although she does not provide us with any hint in that direction, we still can be inspired by her proposal to investigate whether observing/measuring can be viewed as a free process.

Despite the inherent difficulties in defining a collapse in quantum mechanics, we still can trace some general approximation lines. Even with all disputes, we might conceive the collapse as an instantaneous event occurring entirely instantly. It seems that the collapse is some \textit{occurrence} ---or in \citeauthor{seibt2004free}'s \citeyearpar[p.~24]{seibt2004free} words, ``a mode of occurrence''. From this, we can examine how her ontology might be used to address the issue from two different, although complementary, directions ---space and time. \citet{seibt2004free} provides a typology of free processes according to their related status to space and time. Here we just put the results obtained by \citet{seibt2004free} referring to what she called ``type 5 processes'':

\begin{quote}
    Type 5 processes are spatially and temporally maximally self-contained. Prime examples of such processes are entities conceived of as masses or stuffs proper. Once water is not viewed as a mixture but as stuff, it is spatially maximally self-contained, and, since it endures, it is temporally maximally self-contained. Taking a `substantivist' view on space and time \textelp{}, space, time, and spacetime are also type 5 processes. \citep[p.~43]{seibt2004free}.
\end{quote}
    
\textit{Self-contained} describes the relation between an entity $E$ totally contained in a region of space-time $s$ in which $E$ occurs and the temporal parts of $s$ where $E$ might occur. Three options are available: 

\begin{enumerate}[i.]
    \item $E$ occurs at all parts of $s$ (\textit{maximally
self-contained})
    \item $E$ occurs at some parts of $s$ (\textit{self-contained})
    \item $E$ occurs at none part of $s$ (\textit{minimally
self-contained})
\end{enumerate}

A collapse is not an entity like a table or an activity as running; however, the ontology put forth by \citet{seibt2004free} conceives activities as entities as well, which seems to be the case for a collapse. At this point, we gathered the right tools to describe a collapse as a free process. This restriction is intended to deal with entities such as tables and stones well-defined in regions of space and time and cannot be simultaneously at any other location. Such a statement seems to go against the idea that a collapse could be a self-contained entity in a region of space-time. Additionally, she says that ``there are no instantaneous activities'' \citep[p.~485]{Johanna-2009}.

As it seems, we have lost the road toward the free processes ontology proposed by Seibt; she considers, however, activities as just a mode of occurrence ---which seems to be the case of an instantaneous collapse. Given its peculiarity, the notion of a collapse in quantum mechanics does not follow many of our agentive understanding of the world since we cannot say that ``a collapse is occurring''. However, it does not prohibit us from conceiving it as some type of occurrence such as ``the collapse will occur when a measurement is made'' or ``the collapse occurred''.\footnote{~For the \textit{aspectual meaning} of a phrase-noun see \citet[p.~28]{seibt2004free}.}  

From here, based on the results above, we can begin to examine if collapsing can be viewed as a free process of some kind. First, we can examine the case from the perspective of time. \textcolor{black}{Since collapsing is a very fast, nonlocal process}, it has only one temporal part, which implies by reflexivity that \textit{all} (just one) temporal parts of collapsing are still collapsing (reflexivity). This first aspect of free processes and time is defined as a \textit{type of process temporally maximally self-contained}. That is to say, each (one in this case) temporal part of a process is still a process of the same type through time. 
Since collapsing has just one temporal part, which is identical to itself by reflexivity, this process is called trivially homeomerous.
 
Second, it might become disputable from the perspective of space if a collapse is spatially related to just one point in space or more. We sustain that independently from the answer; collapsing is still a free process in Seibt's terms. Her approach is fitting collapsing as spatially maximally self-contained; that is to say, all spatial parts are still the same type. Following her steps, we delineate the notion of collapse as a free process of some kind. This way, although contrary to specific concerns presented by her, we can be inspired by her approach \citep{seibt2004free} to state that a collapse might be defined as a ``spatially and temporally maximally self-contained'' \citep[p.~43]{seibt2004free}. 

At this point, it seems we traced some lines from Seibt's approach toward the measurement problem in QM. However, our main point persists. How can we relate consciousness and the process of measuring? The first option is to consider the notion of consciousness. Consciousness is viewed as an assembly of different mental states related to the world at some level. This phenomenological perspective, shared by Whitehead, is open to Seibt's analysis when dealing with free processes. First, it is enough to list some mental states that operate seemingly to the notion of collapse. Take the case of an observation considered a mental state. As a mental state, observing is paying attention to something.

Take the example of observing the sunset. Observing it at 6 p.m. or 7 p.m. is still observing. In this case, observing is homeomerous, because each part of observing is still observing. In this case, this kind of process is temporally maximally self-contained, which means, as we described above: all temporal parts of observing are still observing. On the other hand, observing the sunset is spatially self-contained and unmarked. For instance, observing the sunset can involve different regions of space, such as the sunset from the coast or the countryside, from Earth or Mars, or even from a TV in the living room. As a result, observing is a process that is temporally maximally self-contained (each temporal part is still observing) while it is spatially self-contained and unmarked (the spatial limits of observing are not well-known). Similarly, many other mental states can be considered free processes of some type.

Ultimately, whether we can formally relate consciousness and collapse is still an open question. However, the developments suggested here were useful for finding a common ground where consciousness and collapse rely on the same ontological framework.

As a result, if we take expressions such as observation, measuring, and so on, we can delineate how close the concept of consciousness is related to the process category. Indeed, \citet[p.~93]{seibt2002quanta} underlines that ``[t]he task of ontology is not only to describe the domain of a theory but to offer an explanatory description whose basic terms we can {`agentively understand'}''. Ontological and epistemological interests cross at this point.

\section{Conclusion}\label{sec:conc}

In this article, we argued that the consciousness collapse interpretation of quantum mechanics might benefit from a Whiteheadian process-based metaphysics. The methodology of such a task is the Toolbox Approach to metaphysics, \textit{viz.} to use the devices developed by analytic metaphysics in order to interpret the scientific endeavor.

Traditionally, such an interpretation is tied up with substance dualism and the mind-body problem. Process-based metaphysics of mind does not suffer from the same fate, so this could be an advantage for the CCCI if understood within the proposed process-based metaphysics framework.

To further develop such a metaphysical framework is a task left for future research in the field of the metaphysics of science. Still, we believe that this proposal is an important first step towards the goal of better understanding our metaphysical options in interpreting quantum mechanics. \textcolor{black}{This is just a small step, and there is still much to be done. But, after all, this is how philosophical research programs run: step by step.}

\section*{Acknowledgements}
Raoni Arroyo acknowledges the support of grant \#2022/15992-8, São Paulo Research Foundation (FAPESP). He also acknowledges that an earlier version of Sections \ref{sec:old} and \ref{sec:new} was published (in Portuguese) in \fullcite{arroyo2023}, and that it benefited from the reading from Jonas Arenhart, Caroline Murr, Décio Krause, and Osvaldo Pessoa Jr. All three authors would like to thank the three anonymous referees for pressing us to better develop important points.

\printbibliography

\end{document}